\documentclass[amssymb,aps,prl,twocolumn]{revtex4-1}

\usepackage{amsmath}
\usepackage{amssymb}
\usepackage{amsfonts}

\usepackage{graphicx}
\usepackage{framed}

\begin{document}

\title{Comment on ``Self-Referenced Coherent Diffraction X-Ray Movie of \AA ngstrom- and Femtosecond-Scale Atomic Motion''}
\author{Kochise Bennett$^{a,b}$}
\email{kcbennet@uci.edu}
\author{Markus Kowalewski$^{a}$}
\author{Shaul Mukamel$^{a,b}$}
\email{smukamel@uci.edu}
\affiliation{$^a$Chemistry Department, University of California, Irvine, California 92697-2025, USA}
\affiliation{$^b$Department of Physics and Astronomy, University of California, Irvine, California 92697-2025, USA}

\date{\today}

\maketitle
In this comment, we challenge the interpretation of ultrafast optical pump X-ray probe diffraction experiments on gas phase $\text{I}_2$ put forth recently by Glownia et al.\ \cite{glownia2016self}.  In that Letter, the x-ray diffraction from a sample perturbatively prepared with excited state population $a$ is given as
\begin{align}\label{eq:Sbuck}
\tilde{S}=N\vert af_e(\mathbf{q})+(1-a)f_g(\mathbf{q})\vert^2
\end{align}
where $N$ is the number of molecules in the gas and $f_{g/e}(\mathbf{q})=\langle g/e\vert\hat{\sigma}(\mathbf{q})\vert g/e\rangle$ is the ground/excited state elastic scattering amplitude (related to the Fourier transform of the electronic charge density $\hat{\sigma}$ operator).
Reference \cite{glownia2016self} assumed ``incoherent mixtures of ground
and excited electronic states", neglecting electronic coherences from the onset. We thus consider only a diagonal electronic density matrix with elements $a$ and $1-a$ and restrict attention to elastic scattering. 
\par
Importantly, the cross term ($f_g(\mathbf{q})f_e(\mathbf{q})$) resulting from the squaring in Eq.\ (\ref{eq:Sbuck}) amounts to heterodyne detection, the interference of a weak signal field ($f_e$) with a strong reference ($f_g$). Such holographic detection has been reported in transient X-ray diffraction in crystals \cite{elsaesser}.  For weak excitations, where only a small fraction of the molecules are excited ($a\ll 1$), the ground-state signal serves as an \textit{in situ} local oscillator for the weaker excited-state signal.  
\par
In Ref.\ \cite{glownia2016self}, it was argued that the linearity in the excitation fraction $a$ of the cross term in Eq.\ (\ref{eq:Sbuck}) renders detection feasible in a heterodyne fashion, while the pure excited-state diffraction scales quadratically in $a$ and is negligible. While we agree that ``This signal is an incoherent sum of the coherent diffraction from each molecule'', we point out that the correct expression \cite{cao1998ultrafast,bratos2002time} for such a signal is
\begin{align}\label{eq:Sone}
S_1=N\langle\hat{\sigma}^*(\mathbf{q})\hat{\sigma}(\mathbf{q})\rangle=N\left(a\vert f_g(\mathbf{q})\vert^2+(1-a)\vert f_e(\mathbf{q})\vert^2\right)
\end{align}
where the expectation value $\langle\dots\rangle=\text{Tr}\left[\dots\rho\right]$, can be evaluated via a trace over the density matrix.   The excited-state diffraction from a gas thus comes linear in the excitation fraction and the amplitude boost from heterodyne detection is neither necessary nor possible.  Equation (\ref{eq:Sone}) and equivalents obtained from the independent atom approximation and rotational averaging have been known in the literature on time-resolved X-ray scattering for many years and appear also in electron diffraction \cite{modxrayphys, cao1998ultrafast,yang2016diffractive}.
\par
The possibility of heterodyne-detected diffraction in crystals (and other systems with long-range order) can be seen by partitioning the total charge density as a sum of molecular charge densities $\hat{\sigma}_\text{gas}=\sum_\alpha\hat{\sigma}_\alpha$ in $S=\langle\hat{\sigma}^*(\mathbf{q})\hat{\sigma}(\mathbf{q})\rangle$.  The diagonal terms in this double-sum generate Eq.\ (\ref{eq:Sone}) while the remaining, two-molecule terms are
\begin{align}\label{eq:Stwo}
S_2=\sum_{\alpha}\sum_{\beta\ne\alpha}e^{i\mathbf{q}\cdot(\mathbf{r}_\alpha-\mathbf{r}_\beta)} \vert af_e(\mathbf{q})+(1-a)f_g(\mathbf{q})\vert^2
\end{align}
where we have assumed identical molecules located at positions $\mathbf{r}_\alpha$.  
This amounts to the observation that the electronic charge densities of distinct molecules are uncorrelated so that, for $\alpha\ne\beta$, we have $\langle\hat{\sigma}_\beta^*(\mathbf{q})\hat{\sigma}_\alpha(\mathbf{q})\rangle =\langle\hat{\sigma}_\beta^*(\mathbf{q})\rangle \langle\hat{\sigma}_\alpha(\mathbf{q})\rangle$. The double-summation pre-factor in Eq.\ (\ref{eq:Stwo}) encodes the long-range structure of the sample and, in crystals, results in bragg peaks at the reciprocal lattice vectors $\mathbf{q}_\text{Bragg}$ \cite{elsaesser}. It is well-established that $S_2$ averages out in the gas phase due to random molecule positions and signals are given by $S_1$ whereas in crystals $S_2$ which scales as $N^2$, dominates $S_1$. Equation (\ref{eq:Sbuck}) is an incorrect crossbreed between the single-molecule and two-molecule contributions to diffraction, with the pre-factor of the former and otherwise the $\mathbf{q}$-dependence of the latter.
Including electronic coherences yields terms that depend on the off-diagonal element of $\hat{\sigma}$ (Eq.\ (10) \cite{bennett2014time}) but Eq.\ (\ref{eq:Sbuck}) never holds, with or without coherence.
\par
In conclusion, the heterodyne single-molecule terms in Eq.\ (\ref{eq:Sbuck}) do not exist and the signal \textit{does} look like that of an ``inhomogeneous gas mixture'' of excited- and ground-state molecules ``where there are no intramolecular cross terms and the intensity distributions simply add''.  The sentence ``The key insight [$\dots$] is that scattering from the excited fraction in each molecule interferes with scattering from its initial state fraction, producing  holographic fringes'' is incorrect.  The molecule does not interfere with itself in the absence of electronic coherence. The observed signal ($\vert f_e(\mathbf{q})\vert^2$) carries no phase information unlike the heterodyne term ($\Re\lbrace f_e(\mathbf{q})f_g(\mathbf{q})\rbrace$).  Moreover, this misinterpretation affects the signal processing as incorrectly dividing by the ground-state charge density magnifies deviations from equilibrium, overestimating the bond elongation of excited I$_2$ (see fig.\ 3 of \cite{glownia2016self}). 
Finally, we note that heterodyne detection is a purely classical effect related to the macroscopic interference of light and has nothing to do with quantum Schroedinger cat states, as was incorrectly stated in Ref.\ \cite{buckpr}.  Quantum features can only be created by electronic coherences, which were neglected in Ref.\ \cite{glownia2016self}.

\acknowledgements
The support of the Chemical Sciences, Geosciences, and Biosciences division, Office of Basic
Energy Sciences, Office of Science, U.S. Department of Energy through award No. DE-
FG02-04ER15571 is gratefully acknowledged. K.B. was supported by DOE.  M.K. gratefully acknowledges support from the Alexander von Humboldt foundation through the Feodor Lynen program.


%

\end{document}